# How different incentives affect homework completion in introductory physics courses

F. J. Kontur and N. B. Terry

This article quantitatively examines the effects that different incentives have on students' homework completion rates in introductory physics courses at the United States Air Force Academy. Our data suggest that no more than 10-15% of the total course points need to be allocated to homework to maximize completion of homework through a grade incentive. In addition, we find that there is a strong relationship between students' perception of the benefits of homework and homework completion rates. Students completed nearly 20% more homework when their average perception of homework increased by 0.5 points on a 4-point Likert scale. Finally, we find that giving students in-class quizzes taken directly from homework resulted in 15% greater homework completion than giving them in-class quizzes that were only conceptually related to homework. We believe that these quantitative findings, many of which agree with common-sense notions of physics educators, will significantly aid designers of introductory physics courses in making homework the most effective learning tool that it can be for their respective courses.

## I. INTRODUCTION

How do you get students to do their homework? Homework is widely used and instructors have devised a number of schemes to motivate students to complete homework problems. Some instructors make students' homework grades a significant percentage of the final course grade. In that case, how much course credit is required? Some instructors do not grade homework at all, instead relying on students' intrinsic motivation to learn the course material. Will this actually work? Some instructors might motivate students by using quiz and/or exams problems which are a close match to the assigned homework problems.

This paper measures homework completion rates in two calculus-based physics courses at the United States Air Force Academy (USAFA) using three different motivational methods: grade incentives, intrinsic motivation, and the use of homework-based quiz problems. The grade incentive, defined as the percentage of the total course grade associated with the homework, varied from 0% to 10% during the semesters studied. Intrinsic motivation was determined using a Likert-style questionnaire which asked students about their perceptions of the value of homework. The impact of intrinsic motivation was determined by examining the correlation between the average perceived learning value of homework for different sections of the course to the average homework completion rates in those sections. The third motivational tool involved giving quizzes which were taken directly from the homework. The homework completion rates associated with these three methods are compared and contrasted.

## II. COURSE OVERVIEW

All USAFA students are required to take two semesters of calculus-based introductory physics in order to graduate. The first-semester course covers Newtonian mechanics concepts. The second-semester course covers electricity and magnetism concepts as well as basic optics. The first-semester course will hereafter be referred to as mechanics, and the second-semester course will hereafter be referred to as E&M.

### A. Student population and demographics

Each semester, the combined enrollment of the two introductory physics courses at USAFA is approximately 1,000 students. As mentioned before, all USAFA students must take the two-semester introductory physics sequence in order to graduate, but many of the students who take these courses will go on to major in subjects outside of science and engineering fields. Based on data for the Class of 2013, 46% of USAFA students major in the humanities and social sciences, 44% major in mathematics, the physical sciences, or an engineering field, and 9% major in biology or geospatial science. The students accepted to USAFA typically have strong academic backgrounds. For the Class of 2013, 82% came from the top quarter of their high school class, 53% came from the top tenth of their high school class, and 9% were either the salutatorian or valedictorian of their high school class. Their average math SAT score is 664, their average math ACT score is 30.3, and their average science reasoning ACT score is 29.4. The male:female ratio for the Class of 2013 was 4:1. Students admitted to USAFA must fall in the 17-22 age range. All students live in dormitories on campus, and class attendance is mandatory.

The standard course sequence for USAFA students places mechanics in the spring of their freshman year and E&M in the fall of their sophomore year. Because of this sequencing, the spring semester is the large-enrollment offering of mechanics, with between 700-900 students enrolled in the course. The fall semester of mechanics is the smaller-enrollment offering, with between 200-300 students enrolled during that semester. By the same rationale, the fall semester is the large-enrollment offering of E&M while the spring semester is the smaller-enrollment offering, with course enrollments being similar to the large and small-enrollment offerings of mechanics

### B. Pedagogy

The introductory physics courses at USAFA are taught in sections where the enrollment is set at approximately 20 students. There are, on average, 14 different instructors for the large-enrollment offerings of the courses and 5 different instructors for the smaller-enrollment offerings. Regardless of instructor, all students in the course have the same textbook, use the same syllabus, complete the same assignments, and take the same quizzes and exams. During the semesters considered in this study, the textbook for the two courses was *Essential University Physics* by Richard Wolfson [1-2]. The learning objectives for each lesson were selected from the learning objectives given at the beginning of the textbook chapters. Students received points for doing pre-class work before every lesson. Pre-class questions were a combination of written work and online questions. The written work was graded in class for completion while the online questions were auto-graded by the computer, sometimes for completion and sometimes for correctness, depending on the semester. All of the pre-class questions were based on example problems and important concepts from the reading. USAFA physics instructors are given extensive training in a variety of interactive teaching techniques, including just-in-time teaching [3], peer instruction [4], think-pair-share [5], and board work problem-solving. This training is done as part of instructors' new faculty orientation. Instructors are highly encouraged to use these interactive teaching techniques in the introductory physics courses.

## C. Homework

For the semesters where credit was given for completing homework, homework assignments were administered through the Mastering Physics online system. The assigned problems were taken from end-of-chapter problems in the textbook, often with the numbers in the problems randomized by the Mastering Physics website. Students were generally assigned 2-3 homework problems for each lesson, with a total of 90-100 problems assigned over the duration of the semester. In most semesters, students were given up to five tries to get the correct answer on homework problems, with no deduction for an incorrect answer until the final attempt. In some semesters, there was a modest deduction in problem score (~3%) for each incorrect answer. Because Mastering Physics allows multiple tries at getting the right answers, most students who attempt to complete the homework receive full credit or close to full credit. So, in the semesters where Mastering Physics was utilized, we use the Mastering Physics homework correctness score as the homework completion score. In semesters where no course credit was awarded for homework, i.e., the semesters when homework was not graded, homework completion was assessed for the purposes of this study by doing periodic random checks of the journals that students were required to purchase for doing their pre-class work and homework.

## III. RESULTS AND DISCUSSION

We analyzed data on three different incentives for completing homework: (1) varying amounts of course credit for homework completion, (2) students' perception homework as a valuable tool for learning, and (3) administering in-class quizzes that were based on homework. The following subsections present data analysis on these incentives.

## A. Awarding Course Credit for Completing Homework

Fig. 1 shows plots of average homework completion (where, as is discussed in the previous section, the correctness score is the same as the homework completion score for the semesters where online homework was utilized) versus overall course credit given for homework completion for 8 semesters of mechanics, 8 semesters of E&M, and combined data for both courses. The data follow a line fit remarkably well in all three plots, with $R^2$ values of 0.77, 0.94, and 0.86. Based on the line fits, an average homework completion of ~80-85% can be achieved by awarding 10% credit in the overall course grade for homework completion. Extrapolation of the line fits suggest that 100% homework completion can be achieved by awarding ~15% of the course credit for homework, though one would expect nonlinear behavior to occur before 100% homework completion takes place. This is consistent with Scharff *et*. *al*. [6], who found that awarding between 10-20% of the course points for online pre-class assignments resulted in maximum completion of those assignments.

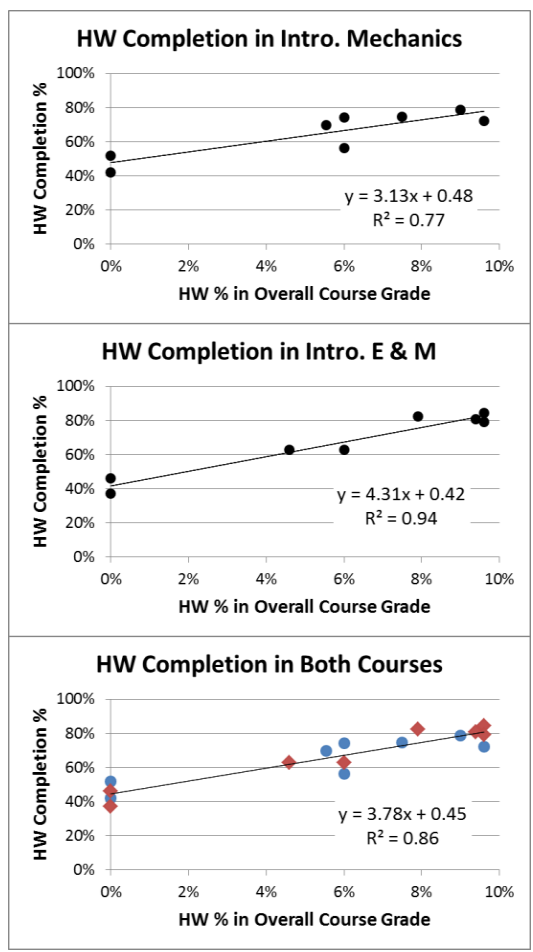

FIG. 1. Plots of homework completion by USAFA physics students as a function of course credit given for homework for 8 semesters of introductory mechanics (top plot), 8 semesters of introductory E&M (middle plot), and combined data for both courses (bottom plot, blue circles are mechanics data and red diamonds are E&M data). Line fits and line fit equations are included in the plots as well.

### B. Student Perception of the Learning Value of Homework

In the 2011/2012 academic year, we stopped using online homework and tried a different method of administering homework. Though homework problems from the book were still assigned to students, those problems were not graded. To determine the impact of having ungraded homework, we asked students to complete anonymous questionnaires at the beginning, middle, and end of the semester. These questionnaires asked about homework and other coursework. For the present study, we examined the results of two of the items on the questionnaire, in which students were asked to indicate their level of agreement with the following statements: (1) Completing homework problems helps me understand the physics concepts, and (2) I tend to perform better on exams when I do all of the homework problems. Students were asked to indicate their agreement or lack thereof with those statements on a 4-point Likert scale: 4-Strongly Agree, 3-Agree, 2-Disagree, 1-Strongly Disagree. Because the questionnaires were completed anonymously, we were not able to match individual student responses to other measures of course performance. However, we did know which course sections the questionnaires came from. This allowed to us to study if the average perception of the value of homework for a course section affected the amount of homework completed by students in that section. These data are shown in Fig. 2. It should be noted that, as was discussed in Section II.B, assignments, quizzes, and exams were the same for all sections of the course, which is likely the main reason why nearly all sections had an average homework value rating that fell in the range of 2.5-3.5 on the 4-point Likert scale. Even over that small range, there are large differences in homework completion. Based on the line fits, a 0.5 point increase in the perceived value of homework will, on average, lead to nearly 20% more homework being completed. This is consistent with previous work [6] which found that the degree to which instructors regular integrated pre-class assignments into the lesson period had a significant correlation with completion rate in cases where 10% or less course credit was awarded for pre-class work. Though the $R^2$ values are much lower for the data sets in Fig. 2 compared to the data sets in Fig. 1, the $p$-values of the data ($p = 0.001$ for the fall 2011 data, $p < 0.0001$ for the spring 2012 data) indicate high statistical significance.

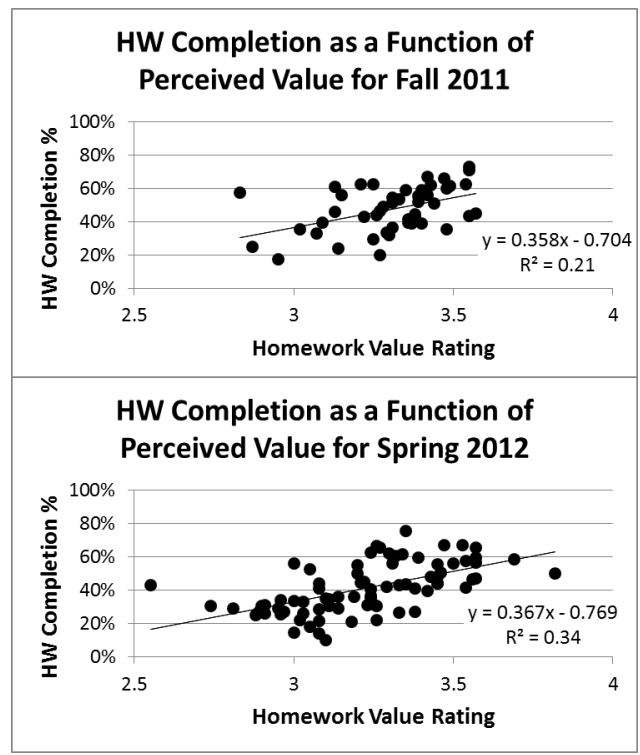

FIG. 2. Plots of average homework completion as a function of average perceived homework value for various sections of the USAFA introductory physics course. Data for the fall 2011 semester are shown in the top plot, and data for the spring 2012 semester are shown in the bottom plot. Homework value ratings were determined by calculating each section's average response on two questionnaire items in which students indicated their perceptions of the usefulness of homework for learning concepts and preparing for exams on a 4-point Likert scale: 4-Strongly Agree, 3-Agree, 2-Disagree, 1-Strongly Disagree.

## C. Relationship between Types of Quizzes and Homework Completion

As was mentioned in the previous subsection, we stopped using graded online homework in the USAFA introductory physics courses during the 2011/2012 academic year. Because homework was not graded, homework completion was incentivized by having 8-10 in-class quizzes that were based on homework. In the fall 2011 semester, the quiz questions were taken directly from homework problems, with only the numbers in the problems being changed. In the spring 2012 semester, the quizzes covered the same concepts as the homework, but were not taken directly from the homework problems. However, students were allowed to use their class journal, including their homework solutions, as a resource when taking quizzes during the spring 2012 semester. In order to obtain homework completion data for this study, random students' class journals were checked to see how many homework problems they had done. After these data were taken, we were able to match students' homework completion scores in fall 2011, when they took mechanics, to their homework completion scores in spring 2012, when they took E&M. Table I shows a summary of the results. We found that, for the 73 students' scores that we were able to compare, on average, homework completion was about 15% greater in fall 2011 mechanics compared to spring 2012 E&M. A *t*-test done on these results shows that these results are statistically significant ($p < 0.0001$) with 95% confidence that the difference in scores lies in the range of 8-22%. As can be seen in Fig. 1, the homework completion rates in mechanics and E&M are, in general, very similar. Considering this, the most likely reason for the difference in the homework completion rates between the two semesters is that the in-class quizzes that used problems taken directly from the homework provided greater incentive for students to do homework than the open-journal quizzes that were conceptually related to the homework but not taken directly from the homework.

TABLE I. Comparison of average homework completion scores for $N = 73$ students whose journals were checked for homework completion in both fall 2011 mechanics and spring 2012 E&M.

| Semester | HW Completion % | Standard Deviation | Std Error of the Mean |
|---|---|---|---|
| Fall 2011 Mechanics | 52.8% | 27.2% | 3.2% |
| Spring 2012 E&M | 37.9% | 27.6% | 3.2% |

## IV. CONCLUSION

We analyzed the homework completion rates associated with three different motivational methods that were used in the introductory physics courses at USAFA – (1) giving course credit for homework, (2) increasing the perceived learning value of homework, and (3) basing quiz questions on the homework. According to our findings, there is a strong relationship between the amount of course credit given for homework and how much homework students complete, but only up to about 15% of the course grade. There is a weaker, but still statistically significant, relationship between students' perception of the learning value of homework and how much homework they do. Finally, students were found to do 15% more homework when in-class quiz questions were taken directly from homework problems compared to when quiz questions were only conceptually related to homework problems. This is true even though, in the latter case, students were allowed to use their written homework solutions as a resource when taking the quiz.

While these findings might not be surprising to most physics educators, we believe that quantifying the common-sense assumptions that most educators make about homework will enable course designers to more finely-tune the way homework is implemented in their physics courses. For example, if one wants to motivate maximum homework completion by awarding course credit for homework, the results of the present study suggest that there is no need to award more than 15% of the course points for homework completion. If, on the other hand, one prefers that students have an intrinsic rather than an extrinsic motivation for doing homework, the evidence presented in this article shows that significant increases in homework completion can be achieved by persuading students that homework is a valuable tool for learning physics. This might be done by regularly incorporating discussions of homework into lessons.

Most teachers employ some combination of intrinsic and extrinsic motivations to get students to do homework. A teacher employing such a combination might award 5% of the course points for completing homework while devoting several minutes at the end of each lesson to a discussion about how the homework problems will build on what students learned in the lesson and, finally, promising students that at least half of the quiz questions will be taken from the homework. Ultimately, each teacher will develop his or her own method for utilizing and motivating homework completion, but the findings in this article will allow physics teachers to make more informed decisions about how homework is utilized in their courses.

**Notes**

Distribution A, approved for public release, distribution unlimited.

**Acknowledgements**

We would like to thank Dr. Lauren Scharff in the Scholarship of Teaching and Learning Center at the United States Air Force Academy for her support and guidance in this research.